\documentclass[12pt,preprint]{aastex}

\newcommand{\etal}{{\em et al.}}            % Defines et al.

\shorttitle{3C 438 Cluster}
\shortauthors{Kraft \etal}

\begin{document}

\title{The Disturbed 17 keV Cluster Associated with the Radio Galaxy 3C 438}
\author{R. P. Kraft\altaffilmark{1}, W. R. Forman\altaffilmark{1}, M. J. Hardcastle\altaffilmark{2}, C. Jones\altaffilmark{1}, P. E. J. Nulsen\altaffilmark{1}}

\altaffiltext{1}{Harvard/Smithsonian Center for Astrophysics, 60 Garden St., MS-67, Cambridge, MA 02138}
\altaffiltext{2}{University of Hertfordshire, School of Physics, Astronomy, and Mathematics, Hatfield AL10 9AB, UK}

\begin{abstract}

We present results from a {\em Chandra} observation of the cluster
gas associated with the FR II radio galaxy 3C 438.
This radio galaxy is embedded within a massive cluster with
gas temperature $\sim$17 keV and bolometric luminosity of
6$\times$10$^{45}$ ergs s$^{-1}$.
It is unclear if this high temperature represents the
gravitational mass of the cluster, or if this is an already high
($\sim$ 11 keV) temperature cluster that has been
heated transiently.  We detect a surface brightness discontinuity in the gas
that extends $\sim$600 kpc through the cluster.  
The radio galaxy 3C 438 is too small ($\sim$110 kpc across) 
and too weak to have created this large disturbance in the gas.  
The discontinuity must be the result of either an extremely powerful
nuclear outburst or the major merger of two massive clusters.  
If the observed features are the result of
a nuclear outburst, it must be from an earlier epoch of unusually
energetic nuclear activity.  
However, the energy required ($\sim$10$^{63}$ ergs) to move the gas on the observed spatial
scales strongly supports the merger hypothesis.
In either scenario, this is one of
the most extreme events in the local Universe.

\end{abstract}

\keywords{galaxies: individual (3C 438) - X-rays: galaxies: clusters - hydrodynamics - galaxies: jets}

\section{Introduction}

The {\em Chandra} and {\em XMM-Newton} Observatories have profoundly influenced
our knowledge of the gas dynamics in clusters of galaxies.
A wide range of phenomena including cold fronts, shocks, and
cavities have been studied in a large number of objects over a wide
range in redshift due to the unprecedented angular resolution 
and sensitivity of these two observatories.
One of the most unexpected observational results of the {\em Chandra}/{\em XMM-Newton}
era has been the non-detection of large amounts of
cooling gas in cluster cores inferred from observations made
with the previous generation of observatories.
A large fraction of the clusters observed with these
two observatories exhibit a disturbed morphology.
The role that AGN outbursts and mergers play in the
energy budget and temporal evolution of clusters is still a
question of considerable observational and theoretical interest.
The study of the most energetic AGN outbursts and mergers
is of particular interest as these relatively rare events
will have the most dramatic and long-lasting effects on the 
cluster gas (B\^\i rzan \etal~2004).

In this Letter, we report the discovery of a $\sim$17 keV cluster associated with the
radio galaxy 3C 438.  This is the hottest cluster known to date, marginally 
exceeding the temperature of even the well studied `Bullet' cluster 
(Markevitch \etal~2002).  The high temperature exceeds that of 
any known relaxed cluster, and is almost certainly not indicative of its gravitational
mass.  The X-ray morphology of the cluster gas is highly disturbed, with 
an X-ray surface brightness discontinuity that stretches at least
140$''$ (600 kpc) in an arc extending south from the core. 
Whether this unusually high temperature and complex morphology is the result 
of a radio outburst or an on-going major merger is not clear, although
we argue that the morphology suggests the latter.
The linear size of the radio galaxy 3C 438 is small relative to
the surface brightness discontinuity, so the observed features
cannot be the result of the current nuclear activity.  There are some
hints of cavities larger than the lobes of 3C 438
to the east and west of the nucleus, perhaps
suggestive of `ghost' cavities created by an earlier epoch of nuclear
activity.  If these features are the result of a nuclear outburst,
the inferred energy to evacuate a single cavity
would be 3.4$\times$10$^{62}$ ergs, at least factor of a few larger 
than the outbursts seen in nearby clusters such as Hydra A (Wise \etal~2006) and
MS0735+7421 (McNamara \etal~2005).

This paper is organized as follows.  In section 2, we present
the observational details and discuss the cleaning of the
data.  Our results are presented in section 3.  Section 4
contains a discussion of the implications of our results.
Section 5 contains a brief summary and conclusion.
We assume a cosmology with $H_0$=71, $\Omega_M$=0.27, and
$\Omega_{vac}$=0.73 throughout this paper (Spergel \etal~2006).  The observed redshift
($z$=0.290, Hewitt \& Burbidge~1991) of the host galaxy of 3C 438
corresponds to a luminosity distance of 1481.6 Mpc, and one arcsecond
is 4.317 kpc.  All uncertainties are at 90\%
confidence for one parameter of interest
unless otherwise stated, and all coordinates are J2000.
This source lies at relatively low Galactic latitude
($b$=-12.98$^\circ$); absorption by gas in our Galaxy 
($N_H$=2.05$\times$10$^{21}$ cm$^{-2}$) (Dickey \& Lockman~1990)
is included in all spectral fits and count rate to flux conversions.

\section{Data Analysis}

The radio galaxy 3C 438 was observed for 47.5 ks with the ACIS-S
instrument on 27JAN02 (OBSID 3967).  The data were reprocessed to
apply the most up to date gain and CTI correction, and the event
file was filtered to remove events at node boundaries.  Standard
ASCA grade filtering (i.e. event grades 0,2,3,4,6) was applied
to the data.  A lightcurve was created in the 5-10 keV band after 
removal of point sources to search for periods of background flaring.  
Intervals where the background rate was more than 3$\sigma$ above the
mean were removed, leaving $\sim$29.5 ks of good time.
X-ray emission from the cluster does not fill the field of view,
so a distant region of the S3 chip was used for background for spectral
analysis.  The composite background files created from multiple observations
near the galactic poles would not be appropriate for this observation
due to the low galactic latitude and significant absorption.
The radio structure of 3C\,438 has been studied previously at several
frequencies (Hardcastle et al 1997, Treichel et al 2001). For the
current paper we made a 1.5~GHz radio map with 1.5$''$ resolution
from archival VLA data, calibrating and reducing them in the standard
manner using AIPS.

\section{Results}

\begin{figure}
\plotone{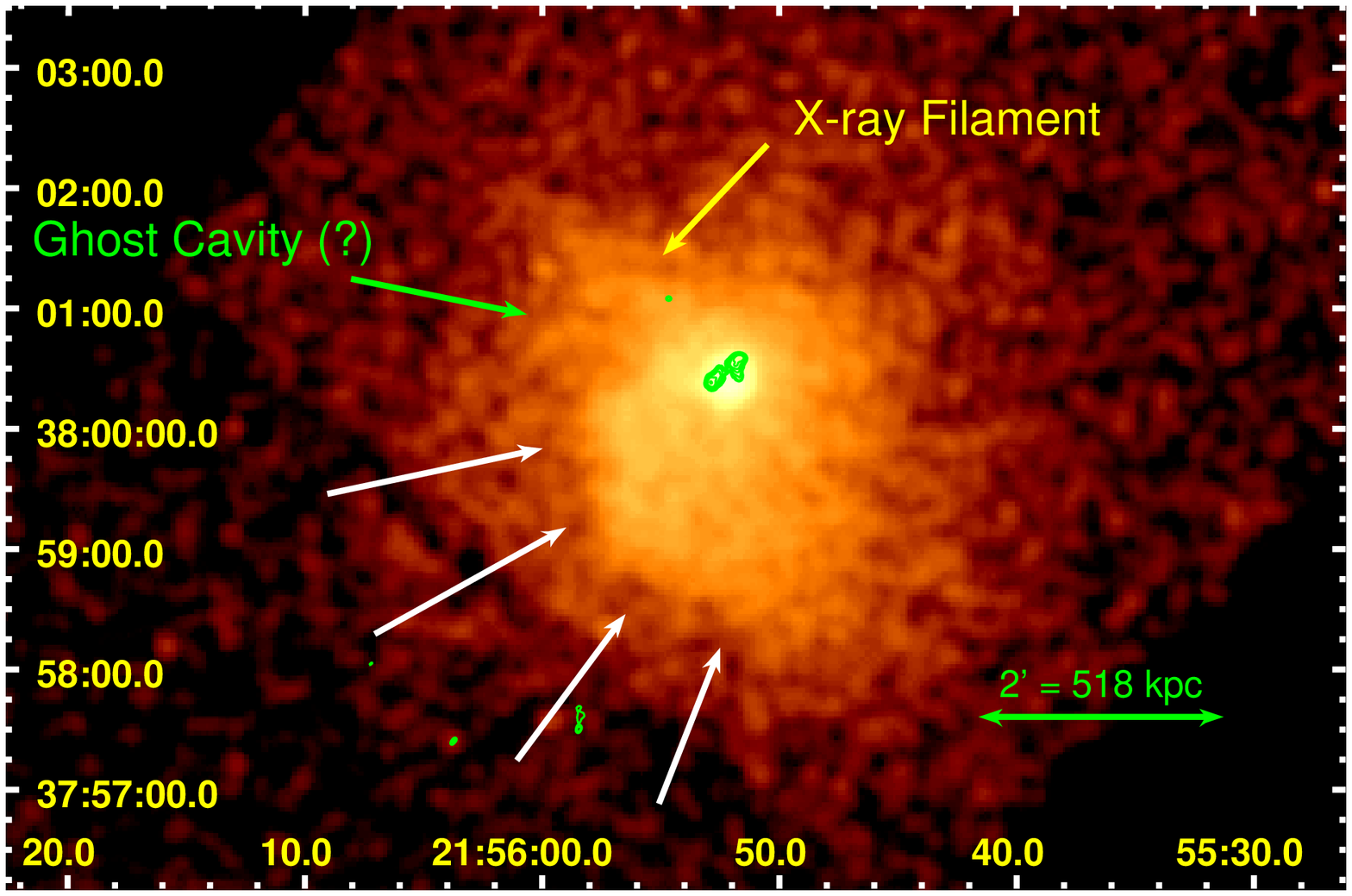}
\caption{Gaussian smoothed ($\sigma$=6$''$) {\em Chandra}/ACIS-S image 
of 3C 438 in the 0.5-5.0 keV band with 1.5~GHz radio contours 
(1.5$''$ resolution) overlaid.  All point sources have been removed.
The radio galaxy lies well within the cluster core, although offset from
the centroid of the X-ray emission.  The white arrows denote the
position of the surface brightness discontinuity, the green arrow the
position of the possible `ghost' cavity, and the yellow arrows the
position of the X-ray filament bounding the northern side of the `ghost'
cavity.}\label{3c438sm}
\end{figure}

\begin{figure}
\plotone{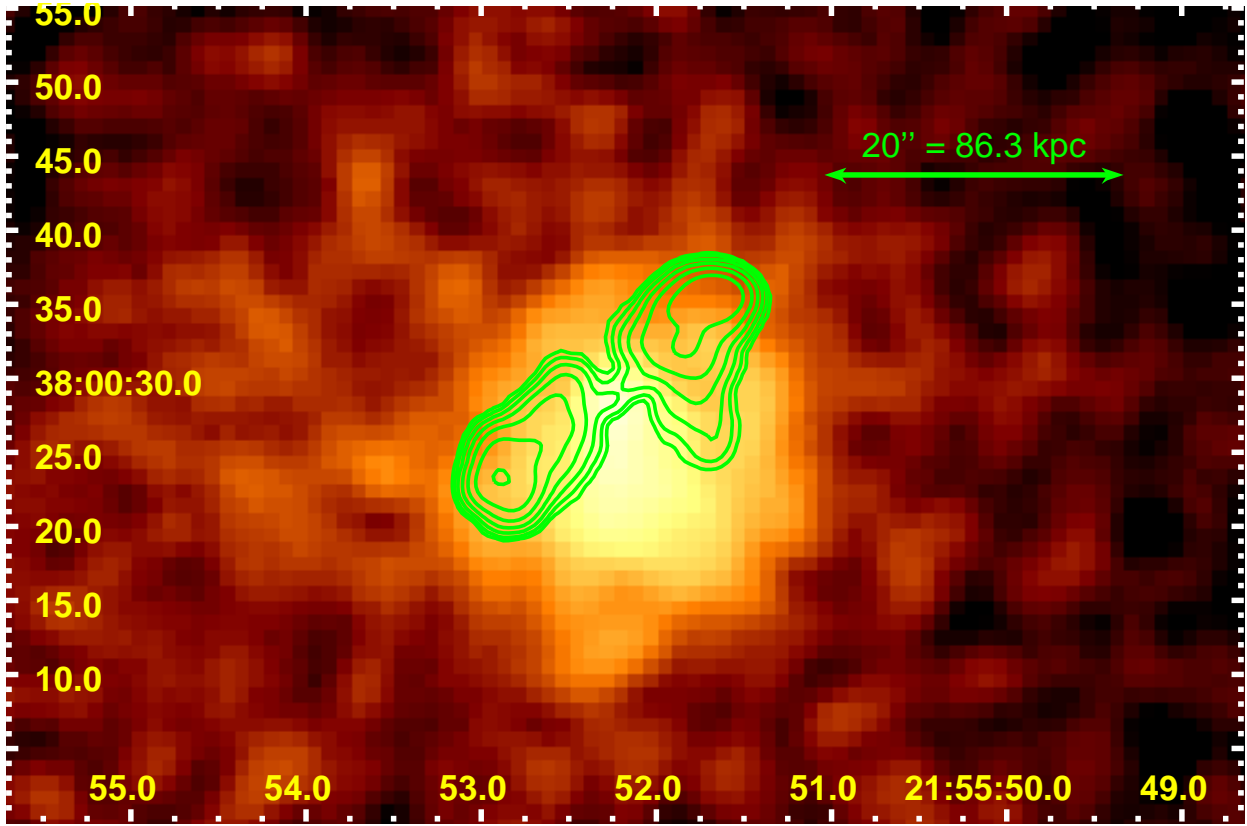}
\caption{Gaussian smoothed ($\sigma$=3$''$) {\em Chandra}/ACIS-S 
image of 3C 438 in the 0.5-5.0 keV band
with 1.5~GHz radio contours (1.5$''$ resolution) overlaid.
The X-ray/radio nucleus of 3C 438 lies $\sim$4.5$''$ (20 kpc) to the NE of 
X-ay peak of the gas, suggesting motion of the galaxy to the NE
in the larger scale cluster emission.}\label{3c438zoom}
\end{figure}

A Gaussian smoothed ($\sigma$=6$''$) {\em Chandra}/ACIS-S image of 3C 438 in 
the 0.5-5.0 keV band with 1.5~GHz radio contours (1.5$''$ resolution) overlaid 
is shown in Figure~\ref{3c438sm}.  
The most striking feature of Figure~\ref{3c438sm} is the surface brightness
discontinuity in the large scale gas (denoted by the white arrows).
This discontinuity is an arc that extends more than 140$''$ (600 kpc).
This suggests that the cluster is highly disturbed on scales of
hundreds of kpc.  There is also a decrement of X-ray counts in a semicircular region
1$'$ east of the host galaxy of 3C 438.  This decrement is suggestive
of `ghost' cavities created by the inflation of radio lobes.
The radio galaxy 3C 438 lies within the X-ray core,
but slightly ($\sim$4.5$''$, 20 kpc) offset to the northeast
from the peak of the X-ray emission.
This is more clearly demonstrated in Figure~\ref{3c438zoom}, a zoom
in on the nuclear region.

\begin{figure}
\plotone{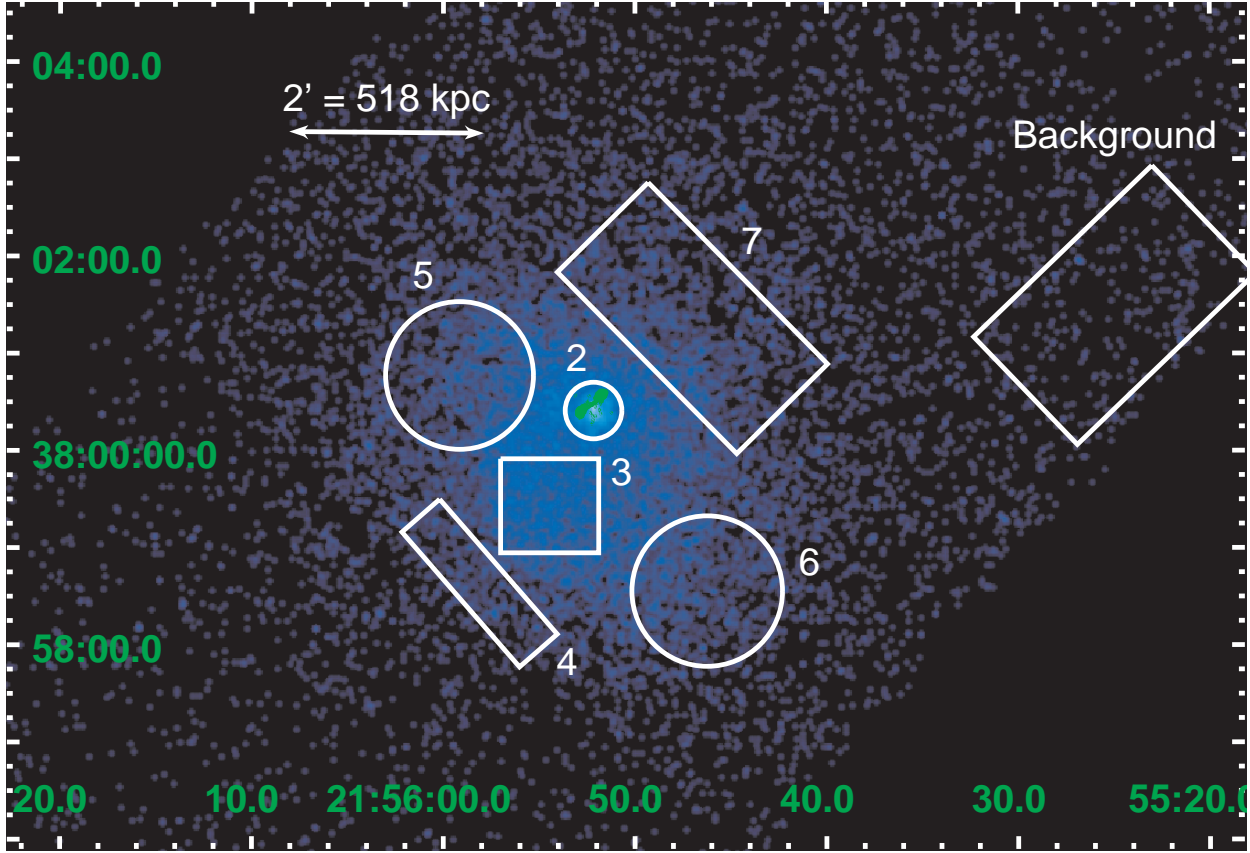}
\caption{Gaussian smoothed ($\sigma$=6$''$) {\em Chandra}/ACIS-S image of 3C 438 
in the 0.5-2.0 keV band with 1.5~GHz radio contours (1.5$''$ resolution) overlaid.
The regions 2 through 7 selected for spectral fitting are overlaid, as well as
the local background region.  Regions 1a (84.5$''$ radius circle centered on the nucleus
of 3C 438) and 1b (an annulus between 84.5$''$ and 138.9$''$ from the nucleus 
of 3C 438) are not shown.}\label{3c438specregs}
\end{figure}

We fitted spectra to eight regions (labeled 1a, 1b, and 2 through 7) 
of the cluster gas.  The positions of regions two through seven and the
distant background region are shown in Figure~\ref{3c438specregs}.
Regions 1a and 1b correspond to a circular region (radius 84.5$''$) 
and an annular region (inner and outer radii of 84.5$''$ and 138.9$''$), 
respectively, centered $\sim$19$''$ SSW of the nucleus of the radio galaxy.
These two regions roughly encompass the X-ray emission inside and
outside of the prominent surface brightness discontinuity.
Single temperature APEC models were fitted with the elemental abundances
frozen to half the Solar value, the column density fixed at the Galactic
value, and the redshift frozen at the value measured for the host
galaxy of 3C 438.  The abundance is poorly constrained if allowed
to vary freely, in part because of the high gas temperature.
The data were binned to a minimum of
50 counts per bin.  The only free parameters were the temperatures and
normalizations.  The results of the fits (all statistically
acceptable) are summarized in Table~\ref{spectab}.

\section{Interpretation}

The best-fit temperature for the gas within 365 kpc of the nucleus of
the cluster (region 1a) is $\sim$17 keV!  As can be seen from Table~\ref{spectab},
the best-fit temperature on larger spatial scales is consistent with this value, 
within large errors.  There are clearly some
cooler subregions contained within 1a, such as the core (region 2)
and the gas on the interior of the surface brightness discontinuity
(region 3).  Even in these cases, however, the {\it cooler} gas
has a temperature of $\sim$11 keV.  
The gas temperature is beyond that of any known relaxed cluster in the local Universe.  
Assuming a uniform temperature of 17 keV, the unabsorbed bolometric luminosity
of the gas within 600 kpc of the host galaxy of 3C 438 is
6$\times$10$^{45}$ ergs s$^{-1}$.
This is therefore one of the hottest, most luminous clusters,
rivaling the distant (z=0.451) cluster RX J1347.5-1145 (Allen \etal~2002, Gitti \etal~2004),
the hottest, most X-ray luminous, reasonably relaxed cluster known.
This temperature is also comparable to
that measured in the highly disturbed `Bullet' cluster ($k_BT\sim$15 keV, with some
regions exceeding 20 keV) (Markevitch \etal~2002).
Extrapolation of the cluster temperature function derived from clusters with z$<$0.1
to 17 keV demonstrates that the space density of such hot clusters is small 
(Markevitch 1998).  There is little or no evolution
of the cluster $L_X - T$ relation out to at least z=0.5 (Mushotzky and Scharf 1997,
Novicki \etal~2002, among many others).
It is therefore unlikely, although not impossible, that the high
gas temperature is representative of the gravitational mass of the cluster.  

\begin{figure}
\plotone{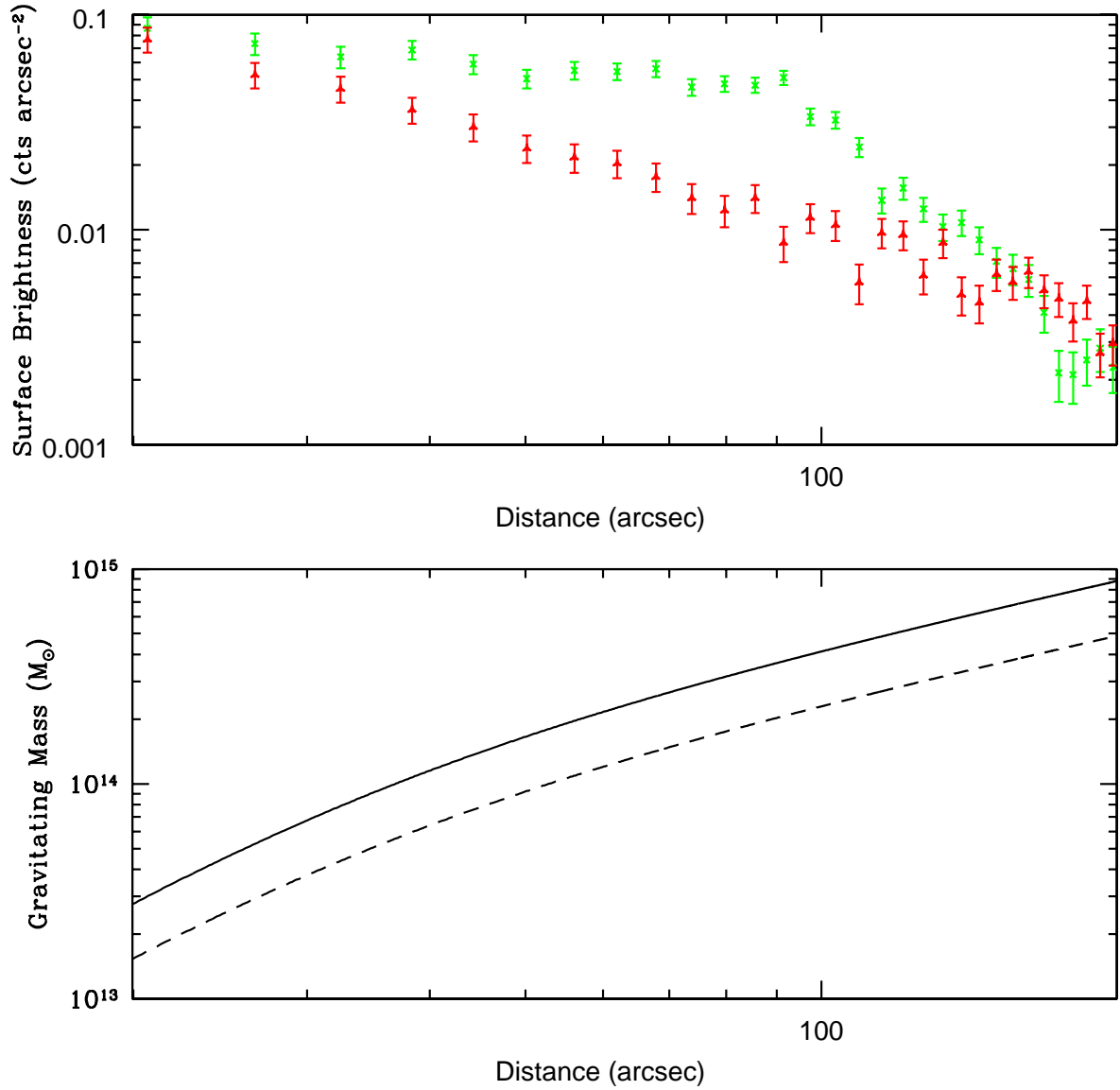}
\caption{The top figure contains the
surface brightness profiles in two 30$^\circ$ wedges to the north (red
triangles) and south (green Xs) of the nucleus in the 0.5-5.0 keV band.  The error bars are
the 1$\sigma$ statistical uncertainty.  The bottom figure shows the
gravitational mass as a function of distance from the nucleus using
the beta model fits for the profile of the northern wedge for two
temperatures (solid line - 17 keV, dashed line - 10 keV).}\label{sbprof}
\end{figure}

To investigate the gravitational mass of the
cluster we consider the surface brightness profile of the cluster.
X-ray surface brightness profiles in two 30$^\circ$ wedges north and south
of the nucleus (centered on the nucleus)
are plotted in Figure~\ref{sbprof} (top).
The surface brightness discontinuity to the south of the nucleus
manifests itself as a sharp change in the slope of the profile
at about 95$''$ (410 kpc) from the nucleus.
The surface brightness of the gas to the north of the nucleus
is relatively smooth, so we fitted a beta-model profile to the
data in this wedge between $20''<r<150''$ ($85<r<650$ kpc)
to determine the undisturbed density profile.
The core radius, $r_c$, is poorly determined in the fits, so we held
it constant at 30$''$ (130 kpc).  We find a best-fit value of $\beta$=0.52$\pm$0.04,
somewhat flat for a cluster of galaxies, but not unreasonable.
The central hydrogen density, $n_0$, is 3.4$\times$10$^{-2}$ cm$^{-3}$ for
$n_e$=1.2$\times n_H$, appropriate for an fully ionized plasma with sub-Solar
elemental abundances.
We have plotted the gravitational mass of the cluster as a function
of distance from the nucleus in Figure~\ref{sbprof} (bottom) using the
parameters derived from the surface brightness profile assuming
hydrostatic equilibrium for two temperatures, 10 and 17 keV.
The extrapolations of these two curves to $r_{500}$ (2.6 and 2.0 Mpc for
17 and 10 keV cluster, respectively) is roughly consistent
with the mass profiles derived from the HIFLUGCS sample of 63 clusters
with $z<$0.1 and 1.0$<$T$<$10 keV (Finoguenov \etal~2001). 
The cluster mass at $r_{500}$ is on the order of a few times
10$^{15}$ M$_\odot$ in either case.
Hydrodynamic simulations of merging clusters show a complex time evolution
of the cluster $L_X$-$T$ relationship (Randall \etal~2002, Rowley \etal~2004), so
that determination of mass in this manner may be biased, although
the variation of $L_X$-$T$ during a merger is smaller than our uncertainties
on the cluster temperature.
In any case, our conclusions below are not sensitive to whether the underlying gravitational
mass profile is representative of a $\sim$10 keV cluster, a $\sim$17 keV cluster, or
an even larger cluster.

The high temperature combined with the highly disturbed morphology suggests that we are
witnessing a dramatic event.  There are at least three possible scenarios
that would explain the high cluster temperature and disturbed morphology.
First, an extremely powerful nuclear outburst
inflated radio lobes in the gas.  Second, this system is a pair of massive 
clusters of roughly equal mass undergoing a major merger.  
Third, the observed X-ray surface brightness
discontinuity is the result of `sloshing' of the gas due to a minor merger.

If the observed features are the result of a radio outburst, it is clear
that the present epoch of nuclear activity cannot be responsible for disturbing
the gas on scales of hundreds of kpc.  
The radio galaxy 3C 438 is too small (with a projected linear
size of only 100 kpc) to be relevant to the observations of the
surface brightness discontinuity. It is also too weak: the amount of
energy required to evacutate the cavities created by the inflation
of the lobes of 3C 438 is only a small fraction of 
that required to displace the gas to create the surface brightness discontinuity.
The bubble enthalpy of the lobes (4$pV$) is $\sim$2.7$\times$10$^{60}$ ergs.
Interestingly, the equipartition enthalpy of the lobes is more than an order of magnitude below
this, implying a substantial departure from equipartition and/or a
significant contribution to pressure from non-radiating particles in
the lobes of 3C 438.  In this scenario in which the large-scale
disturbance in the cluster gas is the result of a nuclear outburst, it must
have been an earlier epoch of AGN activity, although
there is no evidence of any large scale, low frequency radio emission
suggestive of aged, steep-spectrum radio plasma.
There is what appears to be a `ghost cavity' east of the nucleus (spectral region 5) 45.6$''$
in diameter, and perhaps a budding bubble to the W (spectral region 6), similar to what
has been observed in M87 (Forman \etal~2005).
The energy required to completely evacuate the eastern cavity ($pV$ work) is 
3.4$\times$10$^{62}$ ergs assuming that the surface brightness
profile of the northern wedge (discussed above) represents the undisturbed gas.
The enthalpy of two such bubbles is 2.4$\times$10$^{63}$ ergs.
The energy required to create the discontinuity to the west of region 6 
is at least several times this.  
For comparison, the total thermal energy in the gas to this radius
is $\sim$1.5$\times$10$^{64}$ ergs.

There are two problems with this outburst scenario.  First,
and most significantly, the energy requirement, a few$\times$10$^{63}$ ergs, 
strongly argues against this hypothesis.
This would be the most powerful nuclear outburst known, far exceeding
those observed in Hydra A (Wise \etal~2006)
and MS0735+7421 (McNamara \etal~2005), and would require unrealistically
large black hole growth and accretion rate.  Assuming 10\% efficiency, approximately
10$^{10}$ $M_\odot$ would have to be accreted onto the central
SMBH to power an outburst of 2$\times$10$^{63}$ ergs.
In addition, the sound crossing time to reach the end of the discontinuity
is $\sim$2$\times$10$^{8}$ yrs, so the mass accretion rate would
be an unrealistically large $\sim$100 $M_\odot$ yr$^{-1}$.
Second, if the eastern cavity and western bubble are in fact remnants of an earlier epoch
of radio activity, they lie at roughly 45 degree angles to the direction of the
jets of the current outburst.  This would imply that the spin axis of the supermassive
black hole at the center of the host galaxy of 3C 438 has changed direction within
the last 10$^8$ yrs.  This is not entirely implausible as the host galaxy may have merged
with another galaxy in this period.  Optical observations of the host galaxy reveal
that there are several nearby companions, although there is no evidence of
a recent merger (Madrid \etal~2006).  
Recent simulations of the viscous dissipation of buoyant bubbles demonstrate
that their long term evolution may have little relation to the initial
jet axis (Br\"{u}ggen \etal~2005).

The other possibility, and the more likely one in our view, is that we are
witnessing a major merger between two massive clusters.
In this scenario, the X-ray surface brightness discontinuity is a cluster
cold front and the high temperature gas is the shock-heated plasma
of the other cluster.  One cluster is moving roughly to the southeast 
in the plane of the sky.  A temperature of 17 keV implies a relative 
velocity of $\sim$2700 km s$^{-1}$ for the two clusters if this 
temperature is shock heated gas.  This velocity is entirely reasonably, 
and is, in fact, comparable to what has been
inferred for the `Bullet' cluster (Springel \& Farrar~2007).
The primary problem with this interpretation is that
there is only one peak in the X-ray emission, centered
on the host galaxy of 3C\,438.  We might expect to see two
cluster cores if the system is an ongoing massive merger. Due to
the low Galactic latitude of the system, existing optical data are
not good enough to determine whether the distribution of cluster
galaxies is consistent with a major merger scenario.
As stated earlier, the position of 3C 438 and its host galaxy is
slightly offset to the NE from the approximate X-ray peak.
This suggests that the galaxy is moving to the NE and that the
dense core of gas is being displaced from the center of the gravitational
potential by the ram pressure of the larger scale gas.

A third possibility that we consider unlikely is that the gas of the cluster
core is `sloshing' as the result of a minor merger.  
In this scenario, an oscillation of the cluster gas in the gravitational
potential has been induced by a minor (10:1 mass ratio) merger.  The observed
X-ray surface brightness discontinuity is the contact discontinuity 
between two fluids.  Such a model has been invoked to explain surface
brightness edges observed in
Abell 1795 (Markevitch \etal~2001), and there are morphological similarities
between Abell 1795 and the cluster gas around 3C 438.
The energy required to displace the gas in the gravitational potential
of the dark matter to the approximate position
of the discontinuity is $\sim$2.3$\times$10$^{63}$ ergs (1.4$\times$10$^{63}$ ergs
for a 10 keV cluster).
The primary difficulty with the scenario is that the oscillating 
core is unlikely to have dissipated much energy so that the 17+ keV
gas temperature represents the quiescent temperature/mass of the cluster.
In addition, this can only be considered a `minor' merger that only initiates
oscillations of the core gas because of
the extremely high temperature/mass of one of the clusters.
A merger that dissipates $>$10$^{63}$ ergs would be a major event
in virtually any other cluster.

\section{Summary and Conclusions}

We report the discovery of a hot ($>$17 keV), luminous, morphologically
disturbed cluster associated with the radio galaxy 3C 438.
If this temperature is indicative of the gravitating mass of the cluster,
it would be one of the most massive clusters known.  It is most likely
that the high temperature is the result of a massive merger, although 
the limited quality of the data prevents a definitive
conclusion.  Whatever the case, the high
temperature and luminosity of the cluster ensures that the origin
of the disturbance is one of the most energetic events in the local Universe.
Deeper {\em Chandra} and {\em XMM-Newton} observations
are required to further elucidate the dynamics and energetics of this system.
Smaller uncertainties on the gas temperature beyond the discontinuity
would be particularly useful to better constrain the gas dynamics.
Lensing measurements of the gravitational mass of the cluster
would be useful, but the low galactic latitude, crowded 
field, and significant optical extinction may make such 
measurements difficult.

\section{Acknowledgements}

This work was supported by NASA grant NAS8-03060 and the Royal Society.
We would like to thank the anonymous referee for comments that improved
this paper.

\clearpage

\clearpage

\clearpage

\clearpage

\clearpage

\begin{table}
\begin{center}
\begin{tabular}{|c|c|}\hline
Region & Temperature (keV) \\ \hline\hline
1a & 16.6$^{+2.7}_{-2.0}$ \\ \hline
1b & 20.7$^{+10.4}_{-5.2}$ \\ \hline
 2 & 10.7$^{+2.8}_{-1.8}$ \\ \hline
 3 & 10.9$^{+4.0}_{-1.4}$ \\ \hline
 4 & $>$10.0 \\ \hline
 5 & 20.6$^{+19}_{-7}$ \\ \hline
 6 & 17.2$^{+15.2}_{-6.2}$ \\ \hline
 7 & 26.6$^{+27}_{-11.5}$ \\ \hline
\end{tabular}
\end{center}
\caption{Summary of best-fit temperatures and 90\% uncertainties for
several regions of the gas around 3C 438.  Regions 1a and 1b are described
in the text, regions 2 through 7 are shown in Figure~\ref{3c438specregs}.}\label{spectab}
\end{table}


\begin{references}

  \reference{} Allen, S. W., Schmidt, R. W., \& Fabian, A. C. 2002, \mnras, {\bf 335}, 256.

  \reference{} B\^\i rzan, L., \etal~2004, \apj, {\bf 607}, 800.

  \reference{} Br\"{u}ggen, M. Ruszkowski, M., Hallman, E. 2005, \apj, {\bf 630}, 740.

  \reference{} Dickey, J. M., \& Lockman, F. J. 1990, \araa, {\bf 28}, 215.

  \reference{} Finoguenov, A. \etal~2001, A.\& A., {\bf 368}, 749.

  \reference{} Forman, W. R. \etal~2005, \apj, {\bf 635}, 894.

  \reference{} Gitti, M., \& Schindler, S.~2004, A. \& A., {\bf 427}, L9.

  \reference{} Hardcastle, M. J., Alexander, P., Pooley, G. G., Riley, J. M. 1997, \mnras, {\bf 288}, 859.

  \reference{} Hewitt, A., \& Burbidge, G. 1991, \apjs, {\bf 75}, 297.

  \reference{} Madrid, J., \etal~2006, \apjs, {\bf 164}, 307.

  \reference{} Markevitch, M. 1998, \apj, {\bf 504}, 27.

  \reference{} Markevitch, M., Vikhlinin, A., \& Mazotta, P. 2001, \apj, {\bf 562}, L153.

  \reference{} Markevitch, M., Gonzalez, A. H., David, L., Vikhlinin, A., Murray, S., Forman, W., Jones, C., Tucker, W. 2002, \apj, {\bf 567}, L27.

  \reference{} McNamara, B., \etal~2005, Nature, {\bf 433}, 45.

  \reference{} Mushotzky, R. \& Scharf, C. A.~1997, \apj, {\bf 482}, L13

  \reference{} Novicki, M. C., Sornig, M., \& Henry, J. P. 2002, \aj, {\bf 124}, 2413.

  \reference{} Randall, S. W., Sarazin, C. L., \& Ricker, P. M. 2002, \apj, {\bf 577}, 579.

  \reference{} Rowley, D. R., Thomas, P. A., \& Kay, S. T. 2004, \mnras, {\bf 352}, 508.

  \reference{} Spergel, D., \etal~2006, astroph/0603449.

  \reference{} Springel, V. \& Farrar, G.~2007, astroph/0703232.

  \reference{} Treichel, K., \etal~2001, \apj, {\bf 561}, 691.

  \reference{} Wise, M. \etal~2006, astroph-0612100.

\end{references}
\end{document}